\documentclass{jfm}
\usepackage{graphicx}
\usepackage{natbib}

\usepackage{color}

\let\realverbatim=\verbatim
\let\realendverbatim=\endverbatim
\renewcommand\verbatim{\par\addvspace{6pt plus 2pt minus 1pt}\realverbatim}
\renewcommand\endverbatim{\realendverbatim\addvspace{6pt plus 2pt minus 1pt}}
\makeatletter
\newcommand\verbsize{\@setfontsize\verbsize{10}\@xiipt}
\renewcommand\verbatim@font{\verbsize\normalfont\ttfamily}
\makeatother

\ifCUPmtlplainloaded \else
  \checkfont{eurm10}
  \iffontfound
    \IfFileExists{upmath.sty}
      {\typeout{Found AMS Euler Roman fonts on the system,
                   using the 'upmath' package.}%
       \usepackage{upmath}}
      {\typeout{Found AMS Euler Roman fonts on the system, but you
                   don't seem to have the}%
       \typeout{'upmath' package installed. jfm.cls can take advantage
                 of these fonts,if you use 'upmath' package.}%
      }
  \else
  \fi
\fi

\ifCUPmtlplainloaded \else
  \checkfont{msam10}
  \iffontfound
    \IfFileExists{amssymb.sty}
      {\typeout{Found AMS Symbol fonts on the system, using the
                'amssymb' package.}%
       \usepackage{amssymb}%

      }{}
  \fi
\fi

\ifCUPmtlplainloaded \else
  \IfFileExists{amsbsy.sty}
    {\typeout{Found the 'amsbsy' package on the system, using it.}%
     \usepackage{amsbsy}}
    {}
\fi

\newsavebox{\astrutbox}
\sbox{\astrutbox}{\rule[-5pt]{0pt}{20pt}}

\title{Cancellation exponents in helical and non-helical flows}

\author[P. Rodriguez Imazio and P.D. Mininni]{P.\ns R\ls O\ls D\ls R\ls I\ls G\ls U\ls E\ls Z\ns I\ls M\ls A\ls Z\ls I\ls O$^1$ and P.\ns D.\ns M\ls I\ls N\ls I\ls N\ls N\ls I$^{1,2}$}

\affiliation{$^1$ Departamento de F\'{\i}sica, Facultad de Ciencias Exactas y 
                  Naturales, Universidad de Buenos Aires and CONICET, 
                  Buenos Aires, Argentina \\
             $^2$ National Center for Atmospheric Research, P.O. Box 3000, 
                  Boulder, Colorado 80307, USA}

\date{\today}
\pubyear{} 
\volume{000}
\pagerange{\pageref{firstpage}--\pageref{lastpage}}
\doi{S002211200100456X}

\begin{document}

\label{firstpage}
\maketitle


\begin{abstract}
Helicity is a quadratic invariant of the Euler equation in three dimensions. 
As the energy, when present helicity cascades to smaller scales where it 
dissipates. However, the role played by helicity in the energy cascade is 
still unclear. In non-helical flows, the velocity and the vorticity tend to 
align locally creating patches with opposite signs of helicity. Also in helical 
flows helicity changes sign rapidly in space. Not being a positive 
definite quantity, global studies considering its spectral scaling in the 
inertial range are inconclusive, except for cases where one sign of helicity 
is dominant. We use the cancellation exponent to characterize the scaling 
laws followed by helicity fluctuations in numerical simulations of helical 
and non-helical turbulent flows, with different forcing functions and 
spanning a range of Reynolds numbers from $\approx 670$ to $6200$. The 
exponent is a measure of sign-singularity and can be related to the fractal 
dimension as well as to the first order helicity scaling exponent. 
The results are consistent with the geometry of helical structures being 
filamentary. Further analysis indicates that statistical properties of helicity 
fluctuations do not depend on the global helicity of the flow.
\end{abstract}

\section{Introduction}

Theories of fully developed turbulence rely on the energy conservation in inviscid flows to derive the scaling laws of a turbulent system. However, energy is not the only quadratic invariant of the three dimensional (3D) incompressible Euler equation, there is also the helicity
\begin{equation}
H =\int{d^3 x \, \textbf{v}(\textbf{x}) \cdot \textbf{w}(\textbf{x})},
\end{equation}
where $\textbf{v}(\textbf{x},t)$ is the velocity field and $\textbf{w}(\textbf{x},t)=\nabla \times \textbf{v}$ is the vorticity field \citep{Moffatt69}.
Helical motions are observed in a wide variety of geophysical flows, such as supercell convective storms which seem to owe their long life and stability to the helical nature of their circulation \citep{lilly86}. In astrophysics it is well known that helicity plays a key role in the generation of large scale magnetic fields through dynamo action \citep{pouquet76, moffat78}. In hydrodynamic isotropic and homogeneous turbulence, recent studies of flows with net helicity confirmed that there is a joint cascade of energy an helicity to smaller scales \citep{Borue, Chen03, Gomez}, as expected from theoretical arguments \citep{Brissaud}. In the case of non-helical flows, velocity and vorticity tend to align locally creating patches of strong positive and negative helicities. In helical flows, helicity still fluctuates strongly around its mean value, both in space and time, creating locallized patches with helicity much larger or smaller than the mean. As helicity is not positive definite, the study of its turbulent fluctuations is difficult and has received less attention. The scaling laws followed by helical fluctuations are unclear, and few comparisons are available between helical and non-helical flows to understand their impact on the energy cascade. It has been proposed that helical structures may have an impact on the dynamics of turbulent flows as helical regions have the velocity and vorticity aligned, with a resulting quenching of the non-linear term in the Navier-Stokes equation \citep{Tsinober}. Intermittency in the helicity flux has also been recently studied in numerical simulations \citep{Chen203}, and the geometry of helical structures was considered using wavelet decompositions \citep{Farge} and Minkowski functionals in the magnetohydrodynamic case \citep{anvar07}.

In this work we use the the cancellation exponent \citep{Ott92} to study the fast fluctuations of helicity in turbulent helical and non-helical flows. The cancellation exponent was introduced to study fast changes in sign of fields on arbitrarily small scales, and is based on the study of the inertial range scaling of a generalized partition function built on a signed measure. Under some conditions, it was shown to be related to the fractal dimension of structures in the flow and to the Hold\"er exponent \citep{Sreenivasan}. It was used to characterize fluctuations in hydrodynamic turbulence and magnetohydrodynamic dynamos \citep{Ott92}, as well as in two dimensional magnetohydrodynamic turbulence \citep{Pouquet02, Jonathan}. In solar wind observations it was used to show that magnetic helicity is sign singular \citep{Bruno97}. However, to the best of our knowledge, no studies of helical fluctuations in hydrodynamic turbulence have been done using the cancellation exponent. We analyze data stemming from direct numerical simulations (DNSs) of the Navier-Stokes equation in a three dimensional periodic box at large resolutions (up to $1024^3$ grid points) with different external mechanical forcings. Both helical and non-helical flows are considered. We find that helicity fluctuations are sign singular in both cases, and the scaling of helicity fluctuations seems to be independent of the net helicity of the flow. Furthermore, we obtain a relation between the helicity cancellation exponent, the fractal dimension of helical structures and the first order scaling exponent of helicity.

\section{The cancellation exponent}
The cancellation exponent \citep{Ott92} was introduced to characterize the behavior of measures that take both positive and negative values, and to quantify a form of singularity where changes in sign occur on arbitrarily small scales. We can introduce the cancellation exponent for the helicity considering a hierarchy ${Q_{i}(l)}$ of disjoint subsets of size $l$ covering the entire domain occupied by the fluid $Q(L)$ of size $L$. For each scale $l$, a signed measure of helicity is introduced as
\begin{equation}
\mu_{i}(l)=\int_{Q_{i}(l)}{d^3 x \; H(\textbf{x})}\bigg/\int_{Q(L)}{d^3 x \; |H(\textbf{x})|},
\label{eq:mu}
\end{equation}
where $H(\textbf{x}) = \textbf{v}(\textbf{x}) \cdot \textbf{w}(\textbf{x})$ is the helicity density such that the total helicity is $H=\int d^3x H(\textbf{x})$. Since the normalization factor in Eq. (\ref{eq:mu}) is the integral of the absolute value of $H$ over the entire domain, the signed measure is bounded betwen $1$ and $-1$ and can be interpreted as the difference between two probability measures, one for the positive component and  another for the negative component of the helicity density. To study cancellations at a given length scale, we build the partition function
\begin{equation}
\chi(l)=\sum_{Q_{i}(l)}{|\mu_{i}(l)|}.
\label{eq:1}
\end{equation}
In the inertial range, the scaling law followed by the cancellations can be studied by fitting
\begin{equation}
\chi(l)\sim l ^{-\kappa},
\label{eq:chi}
\end{equation}
where $\kappa$ is the cancellation exponent. This exponent represents a quantitative measure of the cancellation efficiency. If such a scaling law exists and its range increases as the inertial range increases, the signed measure is called sign singular. In this case, changes in sign occur everywhere and in any scale considered in the limit of infinite Reynolds number. On the other hand, for a smooth field $\kappa=0$.

\section{Code and numerical simulations}
The data we use for the analysis stems from DNS of the incompressible Navier-Stokes equation with constant mass density,
\begin{equation}
\frac{\partial \textbf{v}}{\partial t}+\textbf{v}\cdot \bigtriangledown \textbf{v}=-\bigtriangledown p+\nu\bigtriangledown^{2}\textbf{v} + \textbf{f},
\label{eq:Navier-Stokes}
\end{equation}
\begin{equation}
\bigtriangledown \cdot \textbf{v}=0.
\label{eq:Incompressibility}
\end{equation}
where $\textbf{v}$ is the velocity field, $p$ is the pressure (divided by the mass density), $\nu$ is the kinematic viscosity, and $\textbf{f}$ represents an external force that drives the turbulence. 

For the analysis, we consider six numerical simulations at different Reynolds numbers and spatial resolutions, with different forcing functions that inject either energy or both energy and helicity. The simulations are listed in Table \ref{tab_Tabla1} and described in more detail in a previous paper \citep{Alexakis06}. Spatial resolutions range from $256^3$ to $1024^3$ grid points, with Reynolds numbers $R_e = UL/\nu$ (with $U$ the rms velocity and $L$ the integral scale) ranging form $\approx 675$ to $\approx 6200$, and Taylor Reynolds numbers $R_e = U\lambda/\nu$ (where $\lambda$ is the Taylor scale) ranging from $\approx 300$ to $\approx 1100$. Two forcing functions were considered: Taylor-Green (TG) forcing, which injects no helicity in the flow (although helicity fluctuations around zero develop as a result of nonlinearities in the Navier-Stokes equation), and Arn'old-Beltrami-Childress (ABC) forcing, which injects maximum (positive) helicity in the flow. The forcings were applied respectively in the shells in Fourier space with wavenumber $k_F=2$ and 3. All runs were continued for over ten turnover times after reaching the turbulent steady state. As a measure of the helicity content of each flow, we computed a relative helicity $\rho_{H} = H/(k_F E)$ (where $E$ is the total energy) averaged in time over the turbulent steady state of each run (see Table \ref{tab_Tabla1}).

\begin{table}
\begin{center}
\begin{minipage}{7cm}
\begin{tabular}{@{}cccccccc@{}}
Run & $N$  & $\textbf{F}$ & $k_{F}$ & $\rho_H$& $\nu$ &$R_{e}$ & $R_{\lambda} $    \\[3pt]
T1 &$256$ & $TG$  & $2$ & $0.03$ & $2 \times 10^{-3}$   &$675$  &  $300$ \\
T2 &$512$ & $TG$  & $2$ & $0.01$ & $1.5 \times 10^{-3}$ &$875$  &  $350$ \\
T3 &$1024$& $TG$  & $2$ & $0.03$ & $3\times 10^{-4}$    &$3950$ &  $800$ \\
A1 &$256$ & $ABC$ & $3$ & $0.79$ & $2 \times 10^{-3}$   &$820$  &  $360$ \\
A2 &$512$ & $ABC$ & $3$ & $0.81$ & $2.5 \times 10^{-4}$ &$2520$ &  $670$ \\
A3 &$1024$& $ABC$ & $3$ & $0.83$ & $2.5 \times 10^{-4}$ &$6200$ &  $1100$
\end{tabular}
\end{minipage}
\end{center}
{\caption[]{Runs used for the analysis. $N$ is the linear grid resolution, $\textbf{F}$ is the forcing, either Taylor-Green (TG) or Arn'old-Beltrami-Childress (ABC), $k_{F}$ is the forcing wave number, $\rho_H$ is the relative helicity, $\nu$ is the kinematic viscosity, $R_{e}$ is the Reynolds number, and $R_{\lambda}$ is the Reynolds number based on the Taylor scale.}
\label{tab_Tabla1}}
\end{table}

\section{Results and analysis}
To study scaling laws of helicity fluctuations in the inertial range of helical and non-helical flows, we performed a signed measure analysis and computed the cancellation exponent $\kappa$ for all runs. Following Eq. (\ref{eq:chi}), this can be done by computing $\chi(l)$ and fitting $\chi(l) \sim l^{-\kappa}$ in the inertial range.

\begin {figure}
\begin{center}
\includegraphics[width=14.0cm]{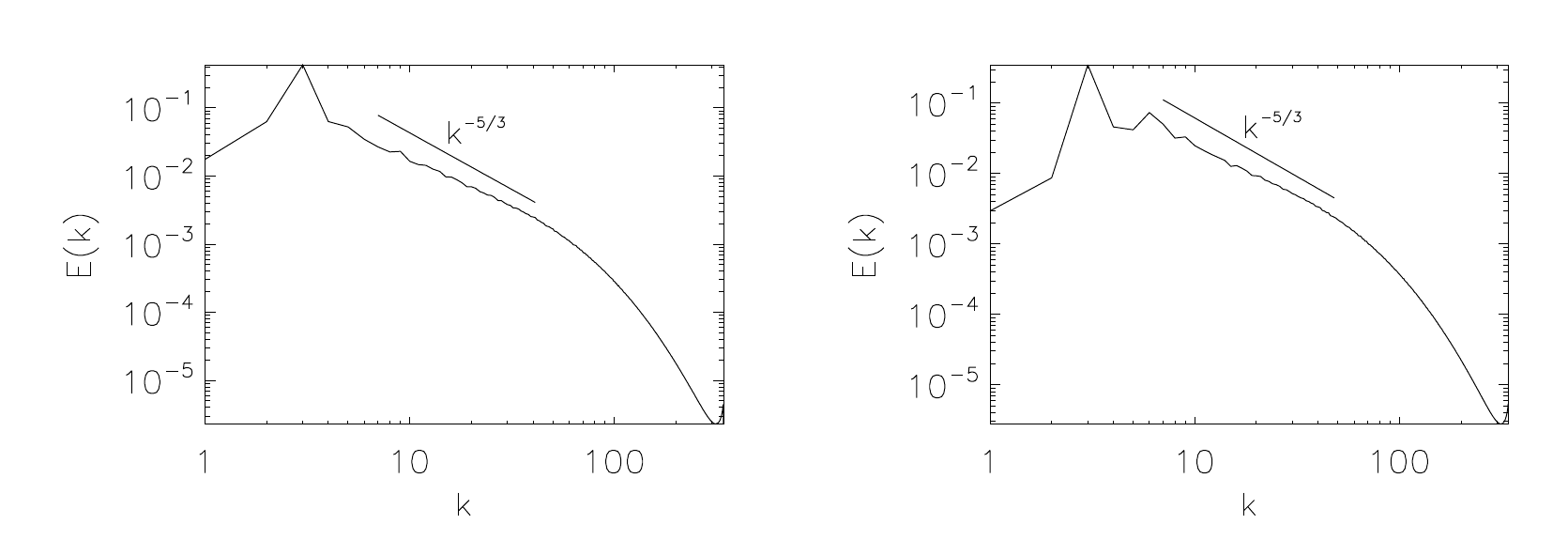}
\caption {Energy spectrum for the A3 (left) and T3 (right) runs. Kolmogorov scaling is indicated as a reference.}
\end{center}
\label{fig:espectros}
\end{figure}

Although a detailed analysis of the simulations can be found in \citet{Alexakis06}, in Fig. \ref{fig:espectros} we show as a reference the energy spectrum of the two simulations at the larger Reynolds numbers attained (runs A3 and T3). The spectra present a short inertial range followed by a bottleneck, as is usual in simulations of hydrodynamic turbulence. The runs also have a range with approximately constant energy flux, as well as a range of scales consistent with Kolmogorov 4/5 law when third-order structure functions are computed. As usual, these ranges have slightly different extensions depending on the quantity studied, and it will be shown that the cancellation exponent shows a scaling range comprised betwen $l\approx 0.08$ and $\approx 0.7$, which correspond roughly to wavenumbers $k\approx 8$ and $\approx 80$, which is wider than but includes the inertial range of both runs. At the lower resolution, the inertial range in the energy spectrum and flux, or in the third-order structure function, is shorter and harder to identify.

\begin {figure}
\begin{center}
\includegraphics[width=6.5cm]{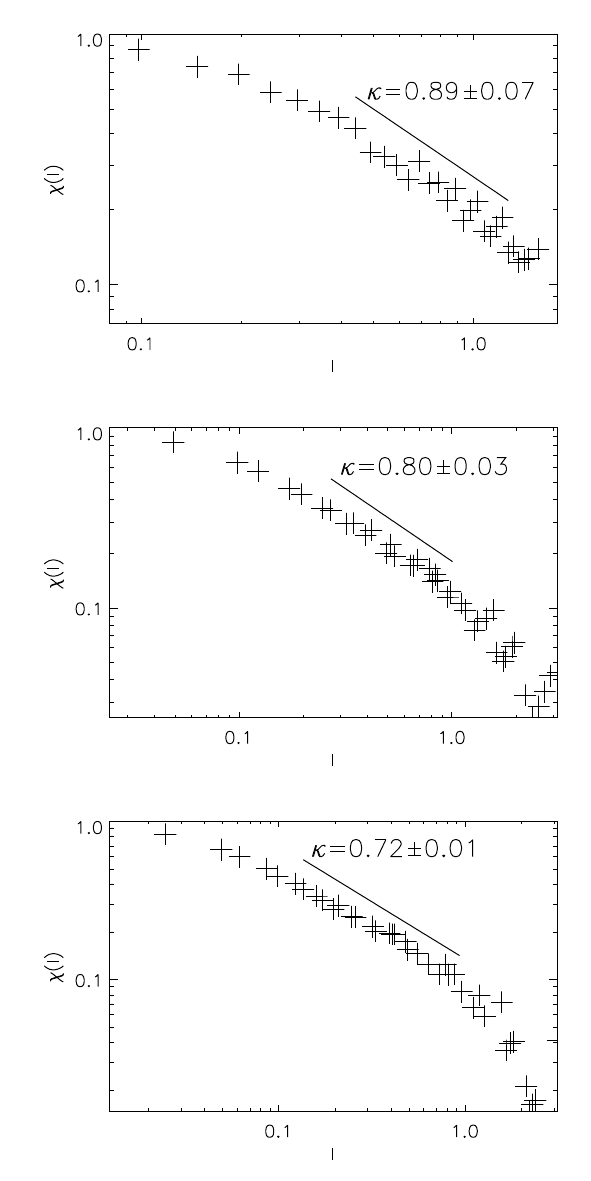}
\includegraphics[width=6.5cm]{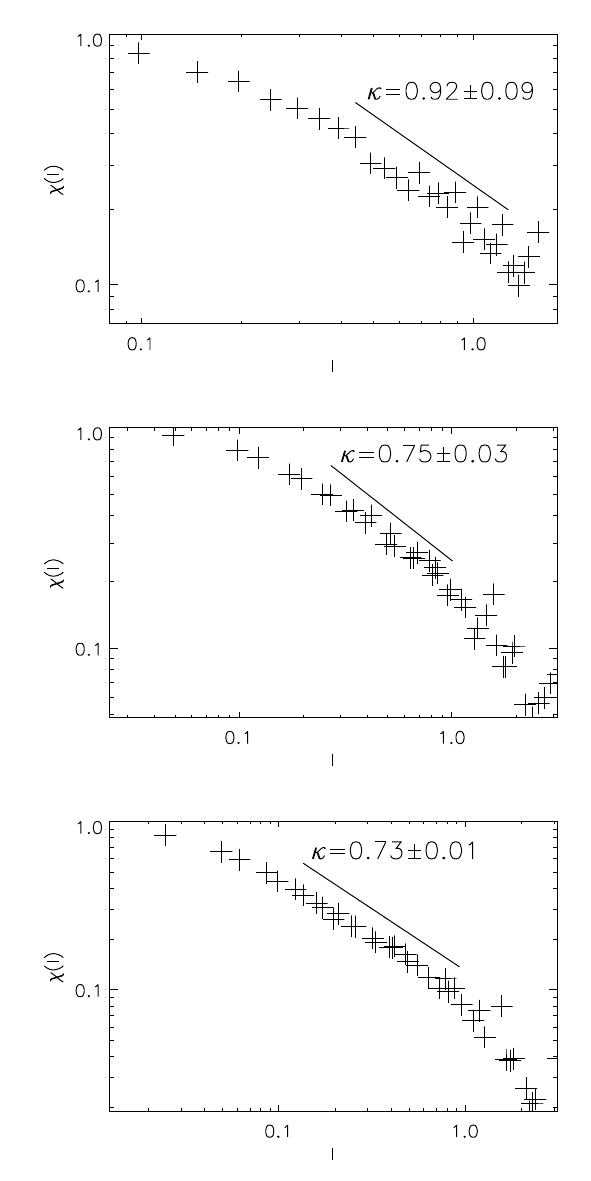}
\end{center}
\caption {Signed measure of helicity as a function of the scale for runs with increasing resolution, from top to bottom: TG forcing (left), and ABC forcing (right) after subtracting the mean value of helicity.  Slopes with the cancellation exponent are given as a reference.}
\label{fig:cancelacionp}
\end{figure}

Fig. \ref{fig:cancelacionp} (left) shows the signed measure of helicity as a function of the scale for runs T1, T2, and T3 from top to bottom (non-helical, with increasing Reynolds numbers). As the Reynolds number increases, a wider scaling range can be identified. The values obtained for the cancellation exponent are $\kappa = 0.92 \pm 0.09$, $0.75 \pm 0.03$, and $0.73\pm0.01$ respectively for runs T1, T2, and T3. The values obtained and the dependence with Reynolds number support the idea that in the limit of vanishing viscosity, helicity fluctuations are sign singular, i.e., that fast changes in sign of helicity occur everywhere in arbitrarily small scales in the flow. This result is consistent with helicity cascading towards small scales, and with observations of the helicity distribution being highly intermittent \citep{Chen203} (albeit the previous studies are only for flows with net helicity).

As mentioned earlier, the flows with ABC forcing are helical. For such flows the direct cascade of helicity and its intermittency has been studied before using spectral quantities. Although helicity fluctuations develop in these flows, given the predominant sign of helicity, fluctuations occur around the mean value and less changes in sign take place. Indeed, if the signed measure of helicity and the cancellation exponent is computed for run A3, we obtain $\kappa = 0.26 \pm 0.04$. This smaller value of the exponent is consistent with the less cancellations, and the larger error is associated to the fact that larger fluctuations are observed in the signed measure as less events whith changes of sign are available. However, we are interested in helicity fluctuations around the mean value, and to correctly consider these fluctuations the mean value of helicity $H$ was subtracted from the helicity density in runs A1, A2, and A3 prior to the analysis. The resulting signed measures are also shown in Fig. \ref{fig:cancelacionp} (right). In this case we also observe scaling of $\chi(l)$ expanding over a wider range of scales as the Reynolds number increases, indicating that also for helical flows helicity fluctuations are sign singular.

The results obtained for the helicity cancellations in helical and non-helical runs are similar. The cancellation exponents are $\kappa = 0.89 \pm 0.07$, $0.80 \pm 0.03$, and $0.72\pm0.01$ respectively for runs A1, A2, and A3. At the highest resolution three ranges can be identified in both the A3 and T3 runs. At the smaller scales, viscous effects dominate and the flow is smooth. This results in less changes of sign in helicity and a shallower distribution of the signed measure. The slope tends to zero, as expected for a smooth flow, and the signed measure tends to one for the smallest resolved scale. At intermediate scales an inertial range is observed, and at larger scales the results are affected by the external forcing. In this scales, large fluctuations in the value of $\chi(l)$ are observed as a result. As mentioned before, the inertial range of $\chi(l)$ is wider than the inertial range observed in the energy spectrum. This is consistent with observations of the helicity spectrum decaying slower than the energy spectrum near the dissipative range in helical turbulence \citep{Alexakis06}. 

\begin {figure}
\begin{center}
\includegraphics[width=6.5cm]{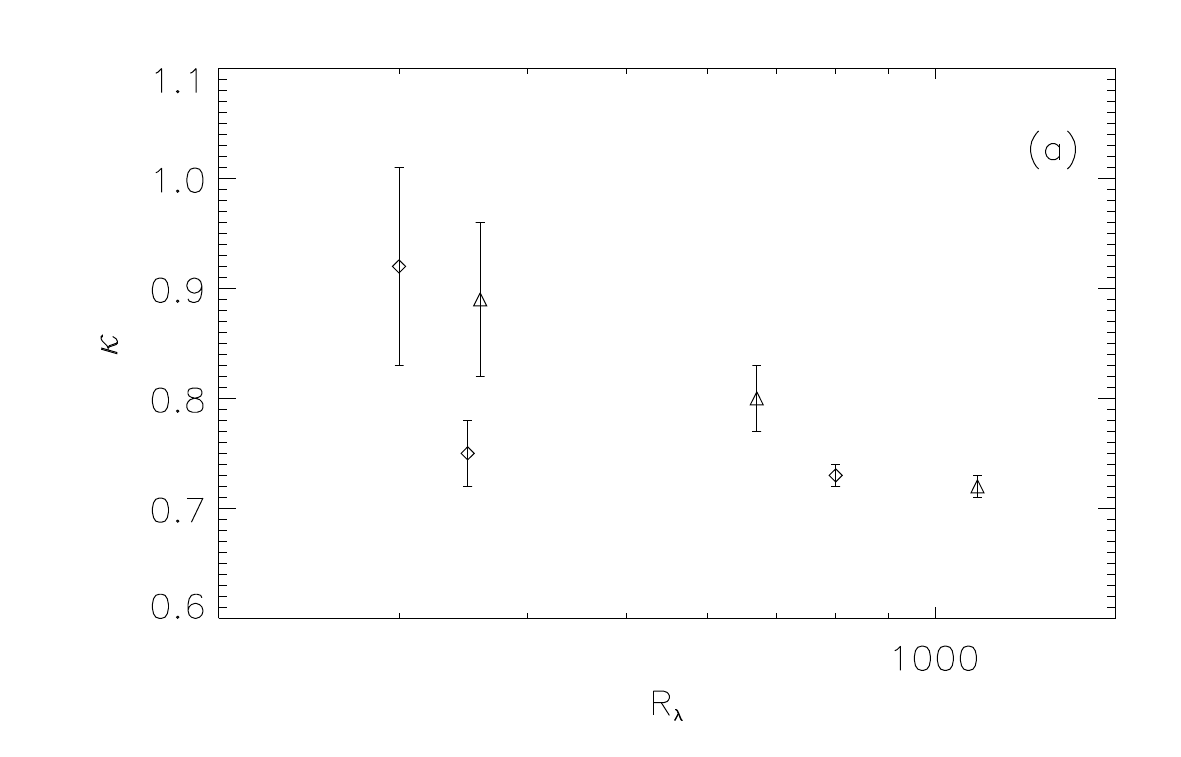}
\includegraphics[width=6.5cm]{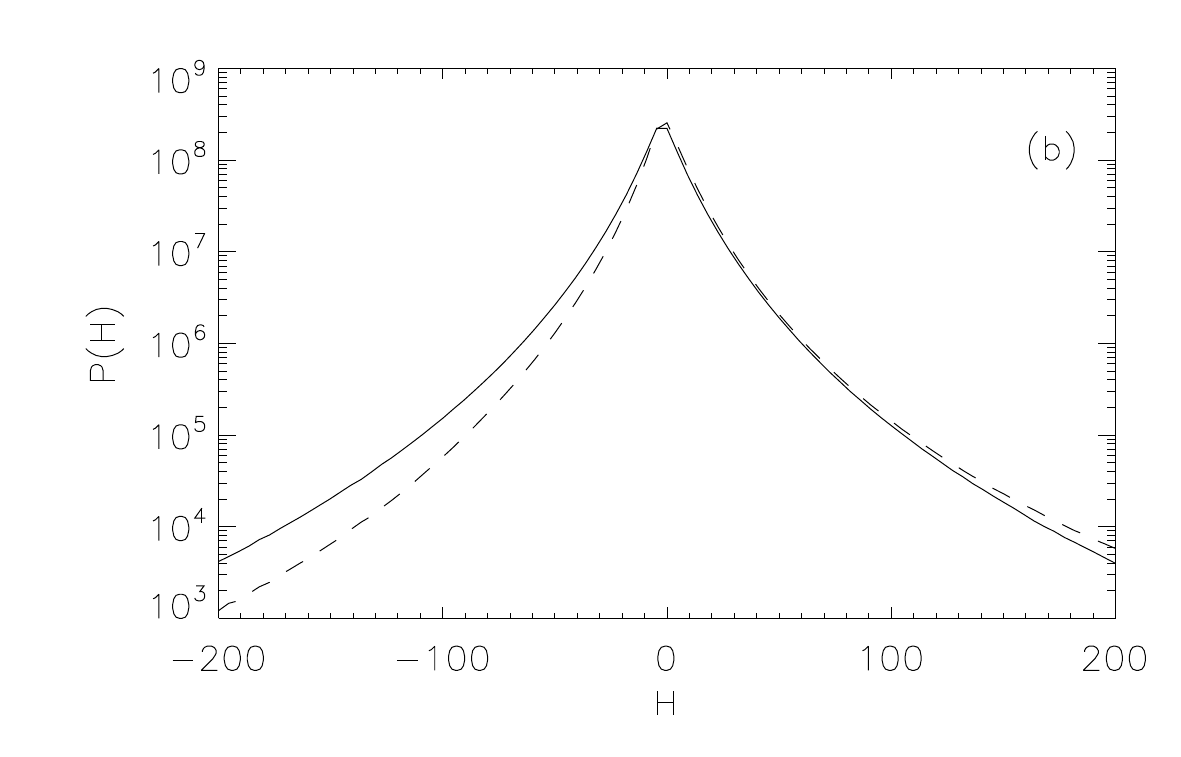}
\end{center}
\caption {(a) Cancellation exponent $\kappa$ as a function of the Taylor based Reynolds numbers for the helical (triangles) and the non-helical (diamonds). (b) Histogram of helicity fluctuations for the T3 (solid) and A3 (dashed) runs.}
\label{fig:kappa}.
\end{figure}

The cancellation exponent obtained in all runs as a function of the Taylor based Reynolds number is shown in Fig. \ref{fig:kappa} (a). For both the helical and non-helical runs $\kappa$ decreases with the Reynolds number, and for the largest Reynolds numbers studied the value of $\kappa$ seems to saturate and is the same within error bars for both helical and non-helical flows. Note that because of the different forcing functions, the helical and non-helical runs have slightly different Reynolds numbers even when the same grid resolution is used. The similarities between the scaling laws of the helical and non-helical flows suggests that statistics of the helicity fluctuations are the same in both cases. This is further confirmed by a histogram of the helicity fluctuations around its mean value for the T3 and A3 runs. As Fig. \ref{fig:kappa} (b) shows, turbulent fluctuations of helicity are similar, with small differences on the tails. 

Helicity fluctuations are sign singular. The fast oscillations in sign point to a highly intermittent quantity that fluctuates rapidly in localized structures at arbitrarily small scales for infinite Reynolds numbers. As mentioned in the introduction, the cancellation exponent can be related to the geometry of such structures, and under some conditions these relations can be rigorously derived \citep{Sreenivasan}. The relations can be found also using simple geometrical arguments. In the following, we extend the analysis of \citet{Pouquet02} for the current density in magnetohydrodynamics to consider the case of kinetic helicity in hydrodynamic turbulence. We assume the helicity is spatially correlated (although not necessarily locally smoth) in $D$ dimensions and uncorrelated in $3-D$ dimensions, where $3$ follows from the dimension of the space. With this choice, a correlated helicity distribution has $D=3$, and a completely uncorrelated helicity distribution has $D=0$. For intermediate cases, we can estimate the signed measure of helicity as follows
\begin{eqnarray}
\chi(l) = \sum_{Q_{i}(l)}{\left|\int_{Q_i(l)}{d^3 x \, H({\bf x})} \, \bigg{/}\nonumber 
    \int_{Q(L)}{d^3 x \, |H({\bf x})|}\right|}\\  
           \sim \frac{1}{L^3\langle H^{2}\rangle^{1/2}}
            \left(\frac{l}{L} \right)^3\left|\int_{Q(l)}{d^3x H(\textbf{x})}\right|, 
\label{eq:fractal}
\end{eqnarray}
where we used homogeneity to replace the sum over all subsets $Q_{i}(l)$ by $(L/l)^3$ (the total number of terms in the sum) times the integral over a generic box $Q(l)$ of size $l$. Since the absolute values in the denominator prevent any cancellation, the integral below was approximated by the characteristic value $L^{3}H_{rms}$ where $H_{rms}=\langle H^{2}\rangle^{1/2}$ is the rms value of the helicity density.

In the inertial range, and still under homogeneity asumptions, the helicity follows some scaling law which can be represented by helicity structure functions $\left\langle \delta H(s)\right\rangle \sim s^{h}$, where $s=|\textbf{s}|$ represents a lineal increment. We can then split the integral over the domain $Q(l)$ in integrals over subdomains with volume $\lambda^3$ such that they separate the contributions from domains where the helicity is correlated of domains where the helicity is uncorrelated in space. Then, the remaining integral in Eq. (\ref{eq:fractal}) can be estimated as \citep{Pouquet02}
\begin{equation}
\left|\int_{Q(l)}{d^3x \; H(\textbf{x})}\right| \sim
\langle H^{2}\rangle^{1/2} \int_{Q(l)} d^Ds \left( \frac{s}{\lambda}\right)^{h}\int_{Q(l)} d^{3-D}s .
\label{eq:contri}
\end{equation}
The uncorrelated dimensions give a value proportional to the square root of their volume $(l/ \lambda)^{(3-D)/2}$. The correlated dimensions give a contribution proportional to $(l/\lambda)^{h+D}$, wich for $h=0$ (locally smoth helicity) is proporcional to their volume. Replacing this results in Eq. (\ref{eq:fractal}), we get that the signed measure of helicity $\chi(l)$ scales as
\begin{equation}
\chi(l) \sim l^{-\left(\frac{3-D}{2}-h\right)}. 
\label{eq:final}
\end{equation}

\begin {figure}
\begin{center}
\includegraphics[width=14.0cm]{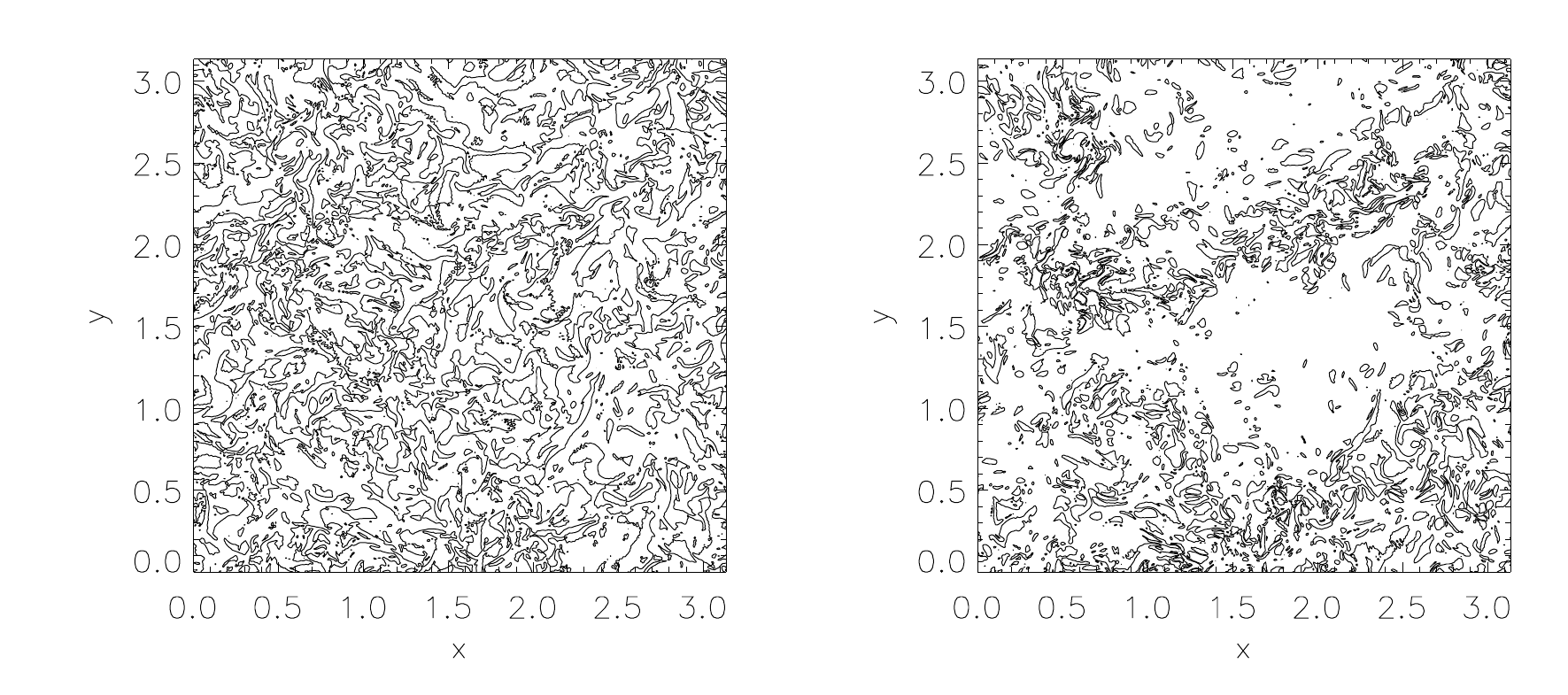}
\end{center}
\caption {Contour levels of helicity density in a slice of the T3 run. Only $1/4$ of the slice is shown. Left: contours with $H({\textbf x})=0$, i.e., places where cancellations take place. Right: contours with $H({\textbf x})=12$. Note how contour levels of zero helicity fill the space, while white patches are observed in the contour levels of non-zero helicity, as well as filamentary structures.}
\label{fig:filamentos}
\end{figure}

Thus the cancellation exponent $\kappa$, the fractal dimension $D$, and the scaling exponent $h$ are related through
\begin{equation}
\kappa=\frac{3-D}{2}-h.
\label{eq:rel}
\end{equation}
A completely smooth field ($D=3$) has $\kappa=h=1$, in agreement with the definition of $\kappa$ and of the first order (H\"older) scaling exponent $h$ for the field (see \citet{Bruno97}). In a turbulent flow, the global helicity is observed to follow Kolmogorov scaling, and therefore we can assume $h=1/3$ \citep{kurien, Alexakis06} . Then from the value of $\kappa=0.73\pm0.01$ it follows that helical structures are filamentary with fractal dimension $D=0.873\pm0.005$, in good agreement with previous observations of helical vortex filaments \citep{Tsinober, Farge, Chen203, Alexakis06}. As an illustration, Fig.\ref{fig:filamentos} shows a slice of the helicity density in the T3 run. Filamentary structures can be identified, specially for the contour levels corresponding to large concentration of helicity. Structures less elongated were confirmed to correspond to filaments that cross the slice perpendicularly. 

\section{Conclusions}
Signed measures of helicity were calculated for turbulent flows with different Reynolds numbers and helicity contents. For both helical and non-helical flows cancellation exponents were calculated, obtaining a value compatible with $\kappa=0.73\pm0.01$ for helicity fluctuations for both cases at the largest Reynolds number considered. The scaling range increases with the Reynolds number, and the value of $\kappa$ obtained indicates helicity is sign singular in the limit of infinite Reynolds number. Simple geometrical arguments were used to link the cancellation exponent $\kappa$ to the fractal dimension $D$ of the structures. To do this we assumed a single scaling exponent $h$ for the lineal helical increments. Assuming $h=1/3$, a value consistent with numerical results, we obtain that helicity fluctuations are filamentary with fractal dimension of $D=0.873\pm0.005$, also in agreement with observations.

\begin{acknowledgments}
The authors acknowledge support from Grants No. UBACYT X468/08 and PICT-2007-02211. PDM acknowledges support from the Carrera del Investigador Cient\'{\i}fico of CONICET.
\end{acknowledgments}


\ifCUPmtlplainloaded \newpage\fi

\end{document}